\documentclass[%
 reprint,
 amsmath,amssymb,
 prb,
 aps,
]{revtex4-1}

\usepackage{graphicx}
\usepackage{dcolumn}
\usepackage{bm}
\usepackage{overpic}
\usepackage[euler]{textgreek}
\usepackage{xcolor}
\usepackage{hyperref}

\begin{document}

\preprint{APS/123-QED}

\title{Optimal power and efficiency of single quantum dot heat engines: theory and experiment}

\author{Martin Josefsson}
\author{Artis Svilans}
\author{Heiner Linke}
\author{Martin Leijnse}
\affiliation{Solid State Physics and NanoLund, Lund University, Box 118, S-221 00 Lund, Sweden}

\date{\today}
\begin{abstract}
Quantum dots (QDs) can serve as near perfect energy filters and are therefore of significant interest for the study of thermoelectric energy conversion close to thermodynamic efficiency limits. Indeed, recent experiments in [Nat. Nano. 13, 920 (2018)] realized a QD heat engine with performance near these limits and in excellent agreement with theoretical predictions. However, these experiments also highlighted a need for more theory to help guide and understand the practical optimization of QD heat engines, in particular regarding the role of tunnel couplings on the performance at maximum power and efficiency for QDs that couple seemingly weakly to electronic reservoirs. Furthermore, these experiments also highlighted the critical role of the external load when optimizing the performance of a QD heat engine in practice. To provide further insight into the operation of these engines we use the Anderson impurity model together with a Master equation approach to perform power and efficiency calculations up to co-tunneling order. This is combined with additional thermoelectric experiments on a QD embedded in a nanowire where the power is measured using two methods. We use the measurements to present an experimental procedure for efficiently finding the external load $R_P$ which should be connected to the engine to optimize power output. Our theoretical estimates of $R_P$ show a good agreement with the experimental results, and we show that second order tunneling processes and non-linear effects have little impact close to maximum power, allowing us to derive a simple analytic expression for $R_P$. In contrast, we find that the electron contribution to the thermoelectric efficiency is significantly reduced by second order tunneling processes, even for rather weak tunnel couplings. 
\end{abstract}
\pacs{Valid PACS appear here}
\maketitle
\section{\label{sec:introduction} INTRODUCTION}
\noindent
Engineering the electronic properties of a nanoscale system using quantum confinement\cite{hicks1993thermoelectric, hicks1993effect, hicks1996experimental, heremans2002thermoelectric, dresselhaus2007new, whitney2014most} or quantum interference\cite{finch2009giant, bergfield2010giant, karlstrom2011increasing, wierzbicki2011influence, trocha2012large, miao2018influence, garner2018comprehensive} is a promising path towards high efficiency thermoelectric energy converters. Quantum dots (QDs) are of significant interest for fundamental research in thermoelectrics because they are ideally zero-dimensional objects with discrete electronic states, and can thus be used as perfect energy filters which only allow electron transport at a single energy. When such a perfect filter is placed between a hot and a cold reservoir (with temperatures $T_h$ and $T_c$), charge and energy flow will be tightly coupled such that every transported charge is associated with a quantized amount of transported energy. This tight coupling is a requirement for reaching the Curzon-Ahlborn efficiency $\eta_{CA}=1-\sqrt{T_c/T_h}$ at maximum power,\cite{curzon1975efficiency, van2005thermodynamic, esposito2009thermoelectric} as well as the Carnot efficiency $\eta_C=1-T_c/T_h$ in the reversible limit.\cite{kedem1965degree, mahan1996best, humphrey2005reversible, van2007carnot,  murphy2008optimal} Because of their potential use in high efficiency devices, the performance of QD heat engines has been studied extensively by theorists.\cite{humphrey2005reversible, murphy2008optimal, esposito2009thermoelectric, nakpathomkun2010thermoelectric, kennes2013efficiency, sothmann2014thermoelectric} However, thermoelectric experiments on QDs are inherently involved as they require a controlled thermal bias over nanometer distances, and measurements of heat flow through nanostructures have so far been limited to a few specific systems and often require some degree of modeling. \cite{jezouin2013quantum, dutta2017thermal, cui2017quantized, mosso2017heat, sivre2018heat} Therefore experiments have mainly focused on measuring the thermally generated open circuit voltage or short circuit current instead of the power and efficienc\cite{staring1993coulomb, dzurak1993observation, dzurak1997thermoelectric, small2003modulation, small2004thermopower, llaguno2004observation, scheibner2005thermopower, pogosov2006coulomb, scheibner2007sequential, svensson2012lineshape, svensson2013nonlinear, thierschmann2015three, svilans2016experiments, svilans2018thermoelectric, prete2019thermoelectric, erdman2019non} (Ref. \onlinecite{jaliel2019experimental} being an exception). Recently, direct power measurements in quantitative agreement with theory predictions, and electronic efficiency estimations were achieved in practice.\cite{josefssonSvilans2018quantum} However, these experiments also highlighted a need for more theory to help guide and understand the optimization of QD heat engines, in particular regarding the conditions under which optimal power and efficiency can be realized in practice. Specifically, when theoretically modeling QD devices it is common practice to assume that the QD is coupled to its reservoirs through very weak tunnel couplings such that the sequential tunneling approximation (SETA) is valid. However, the processes represented by the higher order terms that are neglected in this approximation (co-tunneling, level broadening and renormalization) will in general lead to an increased heat flow, broken tight coupling and lower efficiencies as they introduce an uncertainty in the energy of transported electrons.\cite{turek2002cotunneling, koch2004thermopower} Whether these processes are important for modeling a device will depend on the device's parameters and operating conditions, which has been observed in some experiments.\cite{scheibner2007sequential, josefssonSvilans2018quantum, erdman2019non} Furthermore, the experiments in Ref. \onlinecite{josefssonSvilans2018quantum} also highlighted the critical role of load matching, where an external load is matched to the internal resistance of the device, when optimizing the performance of a QD heat engine in practice. The ability to efficiently predict or determine the optimal load for a given QD is of great practical interest.
\\
\\
In this paper we present thermoelectric measurements on a QD embedded in a semiconductor nanowire together with theoretical calculations that go beyond the SETA. The aim is to understand the conditions for maximum power production of a QD where a single spin-degenerate orbital dominates the electron transport. We derive an explicit, theoretical expression for the external load that optimizes the power output for a given QD using the SETA and linear response, and show that the result agrees well with both more advanced theory, and with a simple procedure for experimentally determining the optimal load. We also show that non-linear and second order effects have little impact on  the optimal load value, but that including these effects is essential for correctly modeling the efficiency. This is because the inclusion of second order effects leads to a spread in the energy of the transported electrons, which will significantly lower the maximum obtainable efficiency even when the tunnel coupling strength is several orders of magnitude smaller than the thermal energy.
\\
\\
The paper is organized as follows. Section \ref{subsec:system} describes the physical principle behind thermoelectric energy harvesting using a single QD. Section \ref{subsec:model} introduces the theoretical model, and \ref{subsec:RTD}  the theoretical framework used for transport calculations. Section \ref{subsec:EXP} describes the experimental device and the measurement setup. The results are found in section \ref{sec:results} where \ref{sec:RP} discusses conditions for maximum power, based on both theoretical and experimental results, and section \ref{subsec:efficiency} focuses on how the theoretical electronic efficiency scales with the device parameters. Finally, section \ref{sec:conclusion} contains some concluding remarks.
\section{\label{sec:theory} MODEL AND METHODS}
\subsection{\label{subsec:system} SYSTEM}
\begin{figure}[hbt!]
	\centering
	\includegraphics[width=0.35\textwidth]{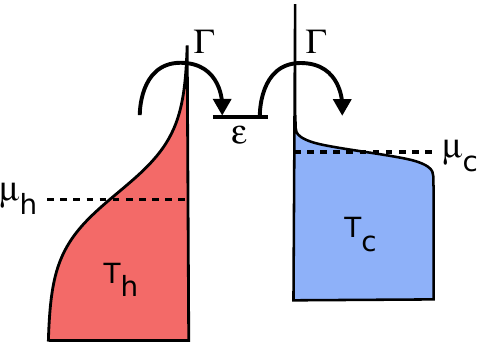}
	\caption{\textbf{Schematic of a QD heat engine.} A QD with a spin-degenerate single particle state at energy $\varepsilon$ coupled to two macroscopic electron reservoirs via electron tunneling (with rate $\Gamma$). The reservoirs are characterized by temperatures $T_{h,c}$ and chemical potentials $\mu_{h,c}$.}		
	\label{fig:Schemtic}
\end{figure}

\noindent
The operating principle of a QD heat engine with one discrete energy level is illustrated in Fig.~\ref{fig:Schemtic}. The QD is tunnel coupled (with coupling strength $\Gamma$) to two macroscopic electron reservoirs at different temperatures $T_{c,h}$. The temperature bias allows the QD to transport electrons up a potential difference when the the electron occupation in the hot reservoir is larger than in the cold at the orbital energy $\varepsilon$, see Fig. \ref{fig:Schemtic}. The device then performs work $\mu_h-\mu_c=eV$ per electron if only sequential tunneling processes ($\propto \Gamma$) are included. The electron transport in the steady-state leads to a stationary current $I$ and generated power $P=-IV$. The power generation is driven by the inflow of heat $J_{Q,h}=I(\varepsilon - \mu_h)/e$ from the hot reservoir to the QD. The resulting thermal-to-electric conversion efficiency is $\eta=P/J_{Q,h}$ if other possible heat losses, such as a phononic heat flow, are neglected. The power generated by the engine is available for consumption in an external serial load $R$. Both $P$ and $\eta$ depend sensitively on the value of $R$ as it self-consistently sets the $V=-RI$ against which the engine must pump the electrons.\cite{josefssonSvilans2018quantum} 
\\
\\
The second order tunneling processes not included in the SETA allow electron transport via virtual intermediate QD states, and when they are included the work performed per electron is no longer exactly $eV$ and a non-zero heat flow $J_{Q,h}$ is possible even in the absence of $I$. Consequently, the existence of the second (or higher) order processes has implications for $P$ and $\eta$.
\subsection{\label{subsec:model} MODEL}
\noindent
We model our QD system using the Anderson model described by the total Hamiltonian
\begin{eqnarray}
	H=H_{D}+\sum_{r=L,R}H_r+\sum_{r=L,R}H_{T,r}.
\end{eqnarray}
The first partial Hamiltonian, $H_D$, describes a QD with a single spin-degenerate energy level where double occupation requires overcoming the Coulomb interaction $U$
\begin{equation}
H_D=\sum_{\sigma=\uparrow\downarrow}n_\sigma\varepsilon_\sigma+Un_{\uparrow}n_{\downarrow}.
\end{equation}
Here $\varepsilon_{\sigma}$ is the single particle energy for an electron with spin $\sigma$. In the absence of a magnetic field the single particle spin states will be degenerate $\varepsilon_\uparrow=\varepsilon_\downarrow=\varepsilon$. The number operator is denoted $n_\sigma=d^\dagger_\sigma d_\sigma$ where $d^\dagger$ ($d$) is the creation (annihilation) operator acting on the QD subspace. $H_r$ describes a non-interacting electronic reservoir 
\begin{equation}
	H_r=\sum_{k,\sigma,r}\varepsilon_{k,\sigma,r}n_{k,\sigma,r},\ \ \ \ \ n_{k,\sigma,r}=c^{\dagger}_{k,\sigma,r}c_{k,\sigma,r}
\end{equation}
where the field operators acting on a reservoir are denoted with the letter $c$. The reservoirs are assumed to be in local thermal equilibrium and can thus be characterized by the Fermi-Dirac distribution $f_r(\varepsilon_r)=[1+e^{(\varepsilon_r-\mu_r)/k_BT_r}]^{-1}$ with chemical potential $\mu_r$ and temperature $T_r$. Any voltage is applied symmetrically to the two reservoirs such that $\mu_{h,c}=\pm\frac{eV}{2}$ and when using the system as a heat engine we choose $T_c=T$ and $T_h=T+\Delta T$. Finally the QD and a reservoir are coupled via the bi-linear hybridization
\begin{equation}
	H_{T,r}=t_{k,\sigma,r}d^\dagger_\sigma c_{k,\sigma,r}+h.c.\ .
\end{equation}
Denoting the amplitude for a tunneling transition $t_{k,\sigma,r}$, we define the bare tunneling rate
\begin{equation}
	\Gamma_{r} = \frac{2\pi\nu_r |t_{k,\sigma,r}|^2}{\hbar},
\end{equation}
where $\nu_r$ is the density of states in reservoir $r$. Furthermore, the wide-band limit is assumed where $\nu_r$ is taken to be constant over an energy range much larger than all other relevant energies in the problem. 
\\
\subsection{\label{subsec:RTD} TRANSPORT THEORY}
\noindent
To calculate the charge and heat currents flowing through the system we use the Real Time Diagrammatic (RTD) theory,\cite{schoeller1994mesoscopic, konig1997cotunneling, leijnse2008kinetic, koller2010density, saptsov2012fermionic, saptsov2014time, gergs2018spin} in which generalized Master equations are set up by expanding the Liouville-von Neumann equation in $H_{T}$. The stationary state reduced density operator of the QD, $\rho_{D}$, is obtained by integrating out the reservoirs and solving the resulting Master equations while imposing probability normalization:
\begin{eqnarray}
	\label{eq:ME1}
	( &\sum_{r=c,h}W_r ) \rho_{D}=0, \\
	&{\text{Tr}}\ \rho_{D}=1.
	\label{eq:ME2}
\end{eqnarray}
Explicit expressions for setting up the kernels $W_r$ to order $\Gamma$ and $\Gamma^2$ are given in the supplemental material of Ref. \onlinecite{gergs2018spin}. The currents flowing through the system can be calculated once $\rho_D$ is obtained. The charge current leaving reservoir $r$ is defined as $\langle I_r\rangle=-e\langle \frac{dN_r}{dt}\rangle$, where $N_r=\sum_{k,\sigma}n_{k,\sigma,r}$, and the energy current as $\langle J_{E,r}\rangle=\langle \frac{dH_r}{dt}\rangle$. The heat current is then obtained from the first law of thermodynamics $\langle J_{Q,r}\rangle=\langle J_{E,r}\rangle-\frac{\mu_r}{e}\langle I_r\rangle$. Following Ref. \onlinecite{saptsov2012fermionic} we use the fact that particle number is conserved in a tunneling process, $[N_r+N,\ H_{T,r}]_-=0$, where $N=n_{\uparrow}+n_{\downarrow}$, to calculate the charge current
\begin{equation}
	\langle I_r\rangle=-i\frac{1}{2}\underset{D}{\text{Tr}}L_{N+}W_r\rho_{D},
	\label{eq:current}
\end{equation}
where $L_{N+}\bullet=[N,\bullet]_+$. A similar treatment is possible for the energy current, however, it requires additional care as $[H_r+H_D,H_{T,r}]_-\ne0$. In the stationary state the energy conservation for electron tunneling instead reads\cite{ludovico2014dynamical}
\begin{equation}
	[H_r+H_D+\sum_{r'\ne r}H_{T,r'},H_{T,r}]_-=0,
\end{equation}
and the energy current can be evaluated as\cite{gergs2017thesis} 
\begin{equation}
	\langle J_{E,r}\rangle = i \underset{D}{\text{Tr}}L_{D+}W_r\rho_D-i\underset{D}{\text{Tr}}W_{T,r}\rho_D,
	\label{eq:J_E}
\end{equation}
with $L_{D+}\bullet=[H_D,\bullet]_+$. The first term on the right hand side of Eq. (\ref{eq:J_E}) is proportional to the eigenenergies of the QD and describes the internal energy change of the QD as a result of electron tunneling. The second term is associated with an energy re-distribution in the barriers and has its leading contribution in $\mathcal{O}(\Gamma^2)$. 
\\
\\
For the QD heat engine to do useful work, an external load must be present. We model the load as a resistance $R$ in series with the QD (see Fig.~\ref{fig:ExpDev}b). Any current generated by the QD thus also flows though the resistance, and the voltage drop $V=RI$ across the resistor acts back on the QD as $-V=(\mu_c-\mu_h)/e$. The current has to self-consistently satisfy $I(V,\Delta T)=-V/R$ if no external voltage is applied across the circuit. The engine then generates power $P=-VI=RI^2$ at efficiency $\eta=P/J_{Q,h}$.
\subsection{\label{subsec:EXP} EXPERIMENTAL SETUP}
\noindent
In this study, we will compare key theoretical results, in particular regarding the optimal load for maximum power, to experimental results. In addition, we also present a technique to efficiently determine the optimal load in experiments. For this purpose we perform power measurements using the experimental setup that is briefly introduced in the following.
\\
\begin{figure}[h!]
	\centering
	\includegraphics[width=0.43\textwidth]{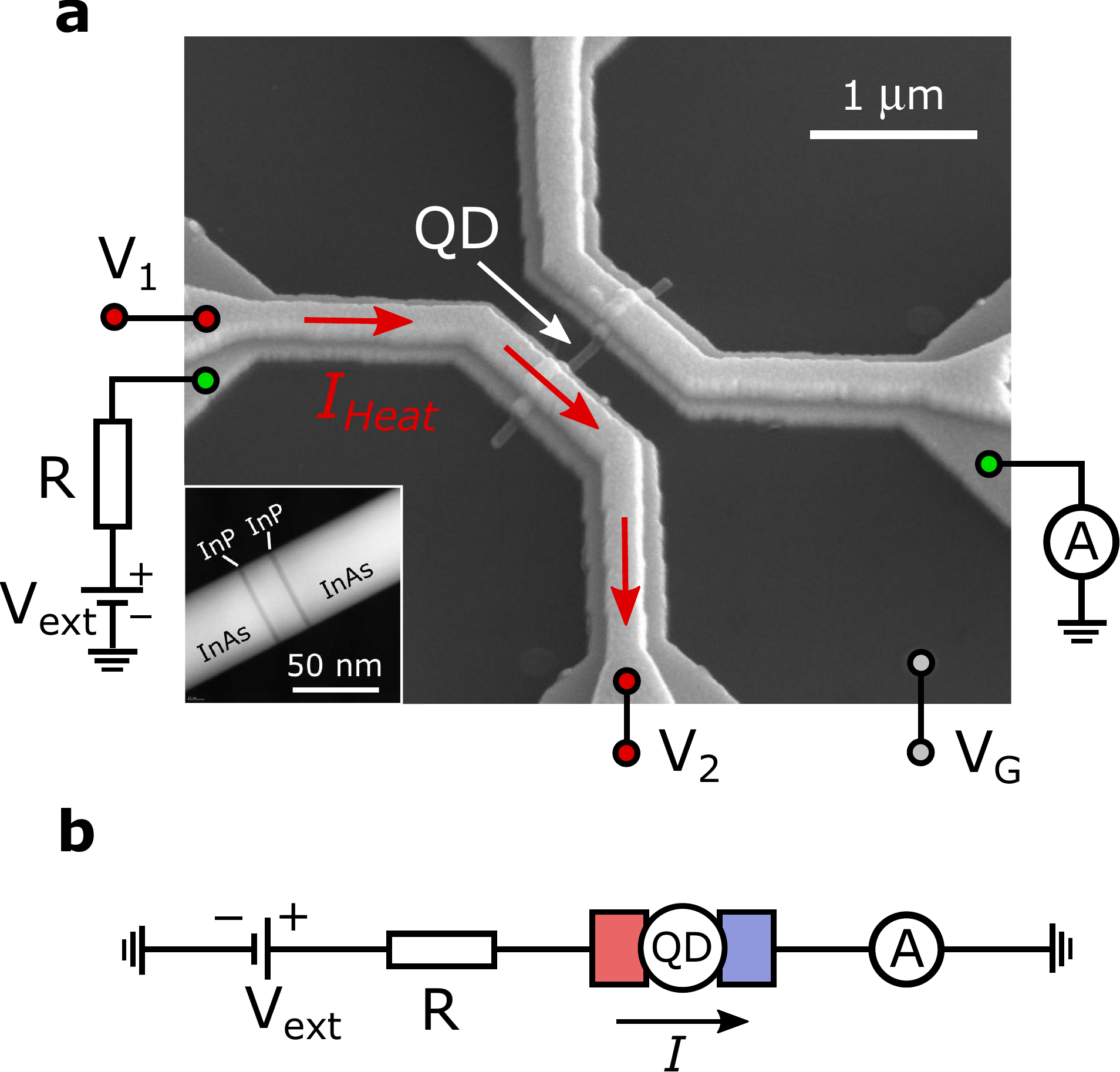}
	\caption{\textbf{Device image and circuit schematics.} \textbf{a} Scanning electron microscope image of a nominally identical device to that used in the experiment. An InAs/InP nanowire is contacted to metallic leads on both sides of the QD (white arrow). A resistance $R$ is in series with the voltage source and the QD. A current $I_{Heat}$ can be run through a top heater lead (electrically isolated from the contact lead) by applying a heater bias $V_{Heat}=V_1 - V_2$. Inset: image of an InAs/InP nanowire (similar to the one used in the device) taken by scanning transmission electron microscope with high angle annular dark field.  \textbf{b} Simplified schematics of the QD heat engine electrical circuit. Components in series from the left: a voltage source applying $V_{ext}$; a resistance $R$, which includes all resistances external to the QD itself, including the preamplifier's resistance and any external load; a thermally biased QD heat engine; a current preamplifier.}	
\label{fig:ExpDev}
\end{figure}
\\
 A scanning electron microscope (SEM) image of a nominally identical device on the same chip is shown in Fig.~\ref{fig:ExpDev}a. The device is fabricated on an n-doped Si substrate covered by a 100 nm thick layer of thermally grown SiO\textsubscript{2}. The device consists of an axially heterostructured InAs/InP nanowire about 60 nm in diameter. An approximately 20 nm long QD segment is defined by two, roughly 4 nm thin, InP segments that function as tunnel-junctions (inset of Fig.~\ref{fig:ExpDev}a). The electrostatic gating of the QD is done by applying a gate voltage $V_G$ to the Si substrate, which is used as a global back gate. The nanowire is contacted to 100 nm thick metallic leads (25 nm Ni and 75 nm Au) for the purpose of applying external electrical bias $V_{ext}$ and measuring current $I$. Additional electrically insulated top heater leads\cite{gluschke2014fully} (25 nm Ni and 100 nm Au) enable application of a sizable thermal bias $\Delta T$ across the QD by running a heating current $I_{Heat}$ through one of the leads and thus dissipating Joule heat.
\\
\\
A simplified schematic of the measurement circuit is shown in Fig.~\ref{fig:ExpDev}b. The QD heat engine is connected in series with a variable resistance $R$ which includes filter resistances of measurement lines (4.5 k\textOmega), input impedance of a current preamplifier (10 k\textOmega) and a variable external resistor. A voltage source can be used to set the external bias $V_{ext}$ across the QD and $R$ when necessary.
\\
\\
When measuring the thermoelectrically produced power of the QD heat engine at a constant $R$, as shown in Fig.~\ref{fig:PVG}a, the variable resistor value is fixed and $V_{ext}$ is set to zero. When instead determining the value of $R$ that maximizes $P$ using an external voltage source the variable resistor is removed and $I$ is measured as a function of $V_{ext}$.
\section{RESULTS}
\label{sec:results}
\noindent
The maximum power and efficiency obtainable by a QD heat engine depend on the quantities $\Gamma$, $R$, $T$, $\Delta T$ and $U$. In the calculations we set $U=100k_BT$, which is on the same order as in the experiments and large enough to allow transport only via one resonance, i.e. the number of excess electrons $N$ fluctuates between configurations that differ by a single electron. The total number of excess electrons on the experimental device is unknown, but it can nevertheless with high accuracy be modeled as having $N=0$ or $1$.\cite{josefssonSvilans2018quantum} The results are divided into two sections. \ref{sec:RP} focuses on the conditions required for the device to produce maximum power, which is discussed in the context of both theoretical and experimental results. Section \ref{subsec:efficiency} theoretically explores what electronic efficiencies should be realizable in real devices.
\subsection{Maximum power}
\label{sec:RP}
\begin{figure}[h!]
	\centering
	\includegraphics[width=0.42\textwidth]{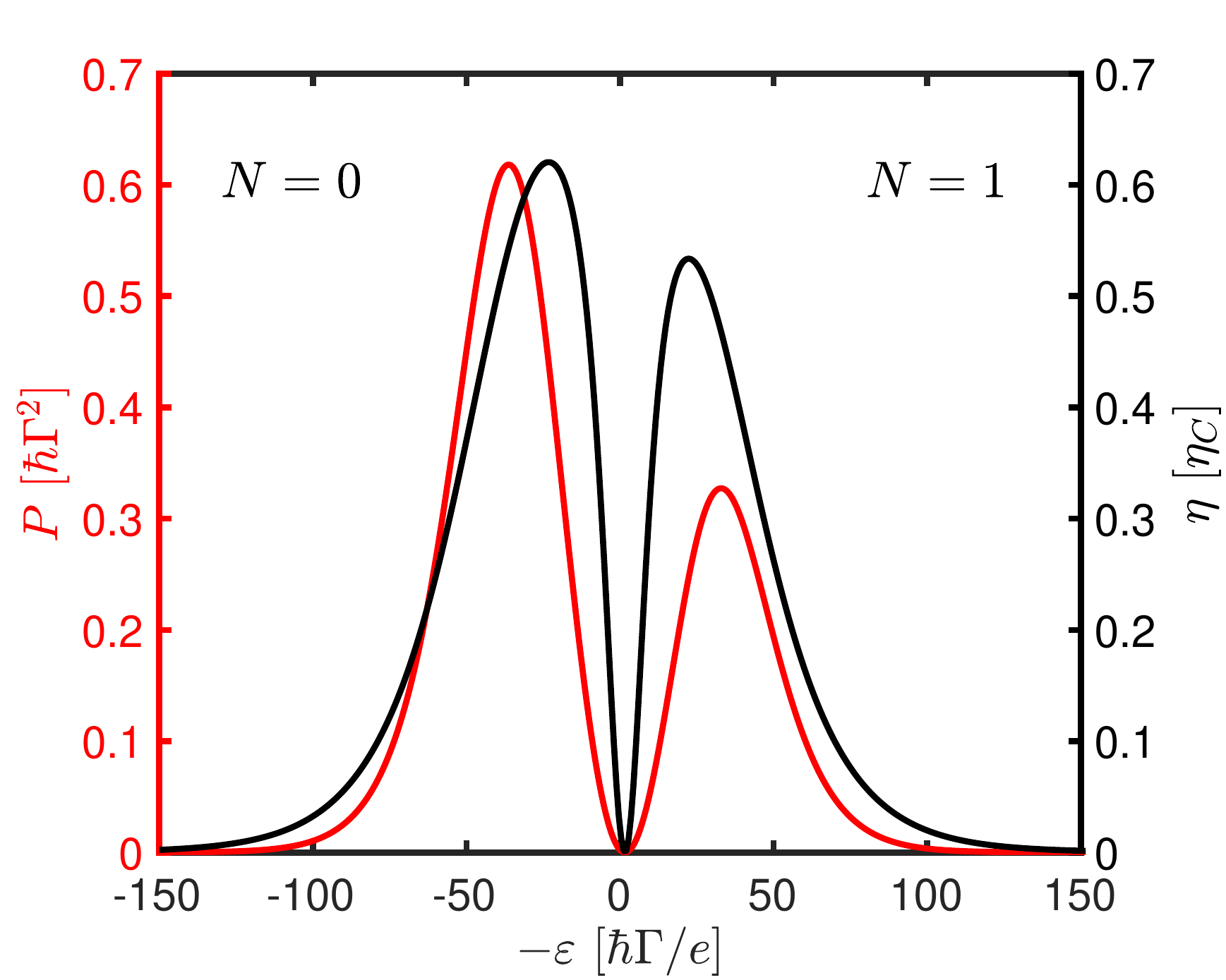}
	\caption{\textbf{Generated power and conversion efficiency.} Generated power $P$ and efficiency $\eta$ as function of level position $\varepsilon$. $P$ and $\eta$ are generally not maximized by the same $V_G$ and there is a trade-off between power and efficiency. $N$ indicate the most likely number of excess electrons on the QD. The results are obtained using $R=40h/e^2$, $T=\Delta T = 10\hbar\Gamma/k_B$ and $\alpha =1$.}		
	\label{fig:PnVg}
\end{figure}

\noindent
Maximizing either $P$ or $\eta$ in a QD heat engine for a given $\Gamma,\ T$ and $\Delta T$ requires optimization of $V$ and $\varepsilon$. Under typical operating conditions $V$ is self-consistently determined by the load $R$ in series with the QD while the resonance energy $\varepsilon$ is set by the gate voltage through $\varepsilon=-e\alpha(V_G-V_G^0)$, where $\alpha$ denotes the lever arm and $V_G^0$ is the voltage for which $\varepsilon=\mu_h=\mu_c$ if $V=0$. Varying $V_G$ enables the QD's orbital to sample different energies of the Fermi-Dirac distributions of the reservoirs, in turn affecting all charge and heat currents flowing through the system. This results in a strong $\varepsilon$-dependence for both $P$ and $\eta$, as can be seen in Fig \ref{fig:PnVg}. Both quantities have a clear overall maximum when there is a high probability to have an even number of electrons on the QD ($\varepsilon>0$). This arises because the state with $N=1$ is spin-degenerate, while $N=0$ is not, resulting in asymmetric currents around the resonance energy ($\varepsilon=0$). See Appendix \ref{app:asym} for a discussion of this effect.
From Fig.~\ref{fig:PnVg} it is also clear that $P$ and $\eta$ do not peak at the same $\varepsilon$, indicating that there is a trade-off between the two. 
\\
\\
Only optimizing $V_G$ is, however, not sufficient to reach the QD's maximal $P$ because $P$ also strongly depends on $R$. Predicting the optimal $R$ for maximum $P$, henceforth called $R_P$, for a real QD heat engine requires, first, knowledge of $\Gamma$, $T$ and $\Delta T$ in the experiment, and second, requires optimizing $P$ in the $V_G$ and $R$ parameter space. Generally, however, it is not trivial to extract these parameter values. Here we instead present a simple procedure for finding $R_P$ experimentally, which is in agreement with the experiments on the same device in Ref. \onlinecite{josefssonSvilans2018quantum} where $R$ was changed manually, and compare these results to the theory. We also show that the principle of impedance matching, where an external load is matched to the internal resistance of the device to maximize power output, is approximately valid in QD heat engines, even far outside the linear response regime.
\subsubsection{Experimental maximum power} 
\begin{figure}
	\includegraphics[width=.42\textwidth]{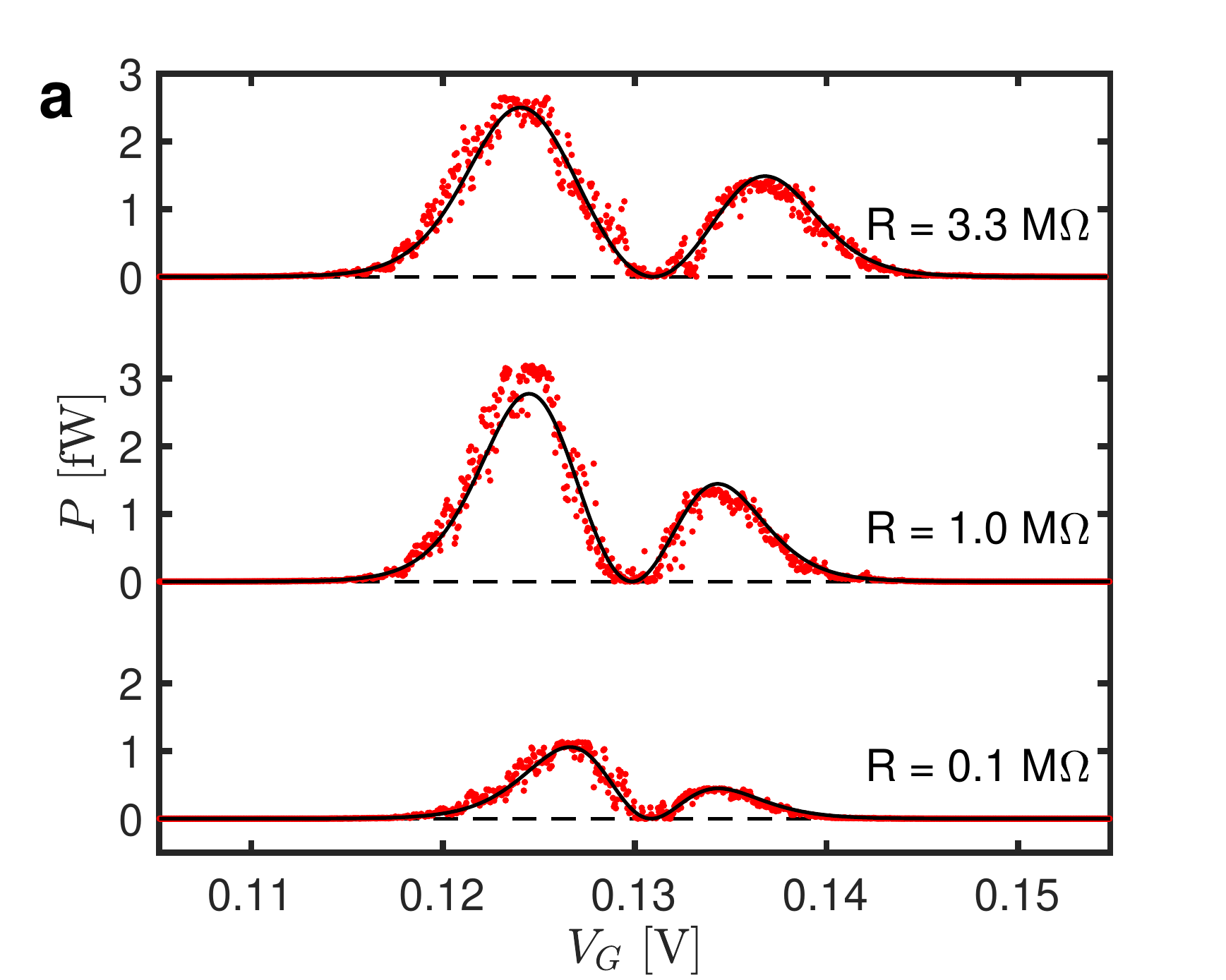}
	\includegraphics[width=.42\textwidth]{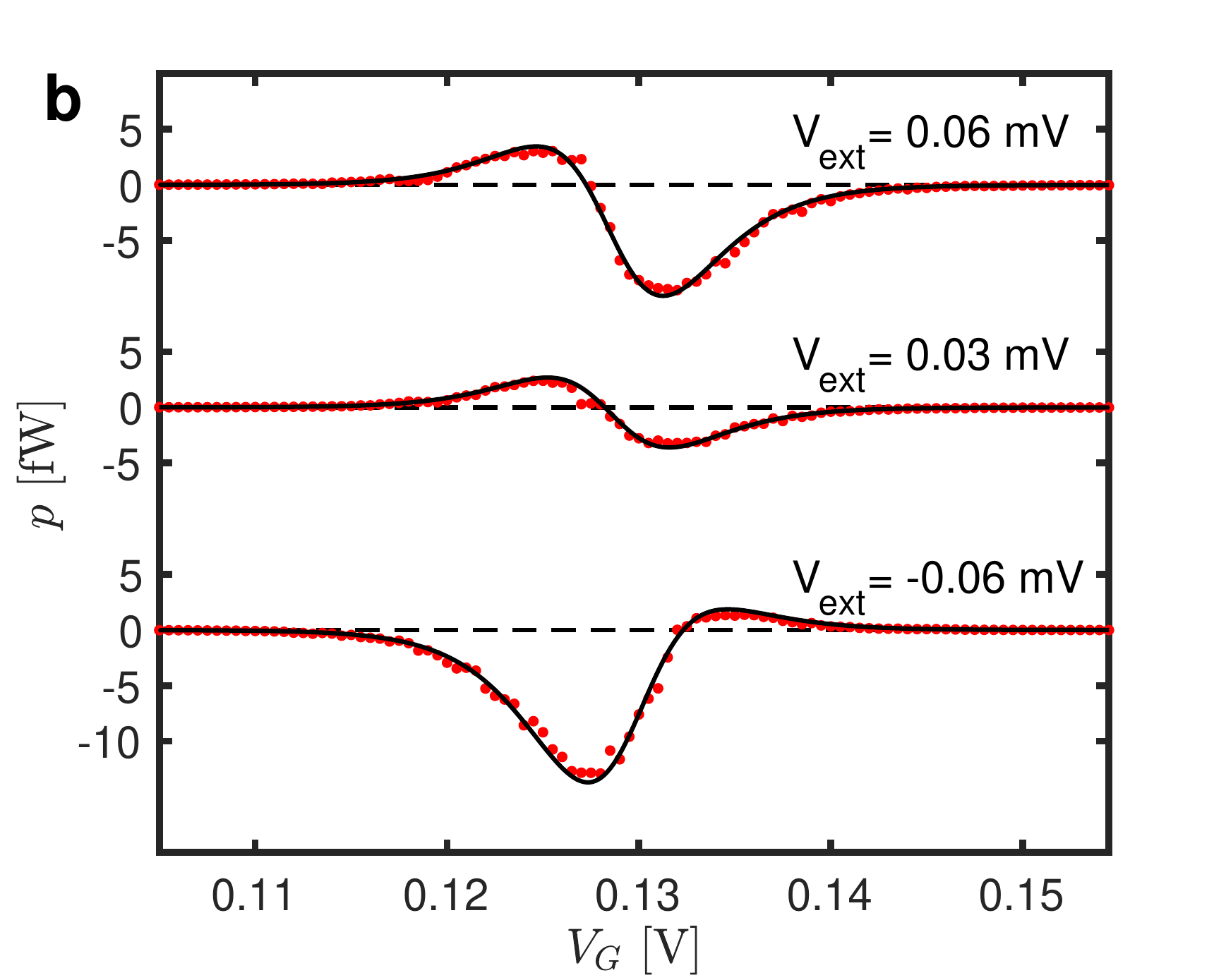}
	\caption[width=1\textwidth]{\textbf{Experimentally measured power.} \textbf{a} $P=RI^2$ generated by the QD and dissipated in $R$ when $V_{ext}=0$. The voltage $V=-IR$ arises due to $I$ flowing through $R$ and self-consistently develops in the circuit. The QD heat engine always works as a generator in this setup ($P>0$). \textbf{b} $p=-IV_{ext}$ measured under a fixed $V_{ext}$ with the external resistor removed. In this setup the QD engine can work either as a generator ($p>0$) or as a heat pump ($p<0$) depending on $V_G$. Red markers represent measured data and solid lines represent calculations based on the second order RTD theory described in section \ref{subsec:RTD}. Both measurements were performed with V$_{Heat}=1000$ mV, which resulted in $T=0.9$~K and $\Delta T=0.6$~K. Experimental device parameters are $\hbar\Gamma=5.8$ \textmu eV, $U=4.9$ meV and $\alpha=0.05$.}
	\label{fig:PVG}
\end{figure}

\begin{figure*}
	\includegraphics[width=1\textwidth]{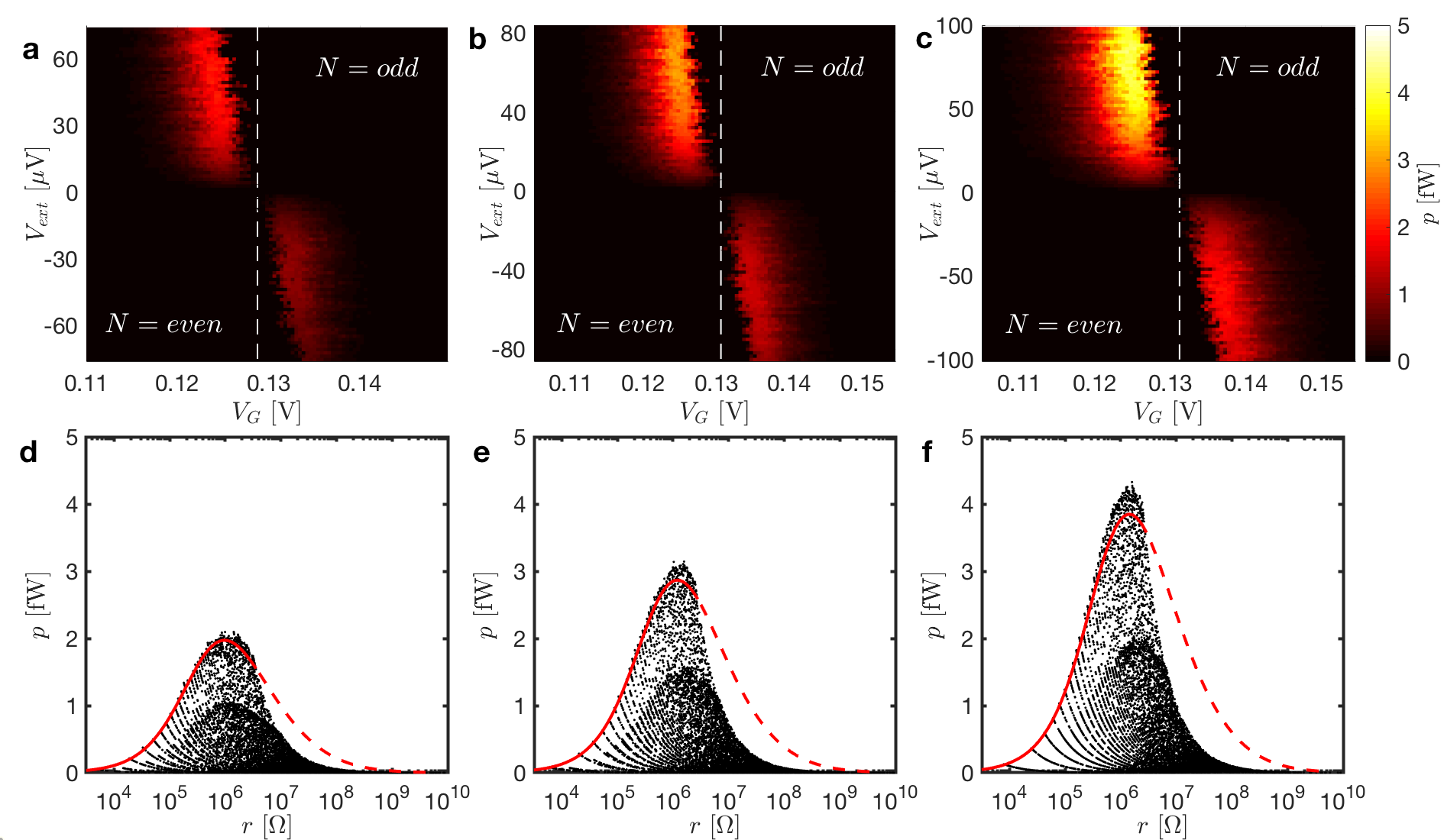}
	\caption[width=1\textwidth]{\textbf{Power production and effective load.} \textbf{a-c} Experimentally measured $p$ as a function of $V_{ext}$ and $V_G$ show two distinct areas of power production corresponding to the large and small peak in Fig. \ref{fig:PnVg} where the QD is most likely in an $N=even$ state for $V_G\lessapprox0.13$ V. Experimental parameters are $\Gamma=5.8$ \textmu eV, $U=4.9$ meV, $\alpha=0.05$ and  $T=0.75$ K, $\Delta T=0.45$ K in \textbf{a}, $T=0.9$ K, $\Delta T=0.6$ K in \textbf{b} and $T=1.15$ K, $\Delta T=0.7$ K in \textbf{c}.  Only regions where the QD produces power are shown in color. \textbf{d--f} Black dots show the data in \textbf{a--c} plotted in coordinates of $p=-IV_{ext}$ and $r=V_{ext}/I$ where each dot corresponds to the $p$ and $r$ for a set of $\{V_G,V_{ext}\}$. The theoretical predictions for maximum power, obtained using the second order RTD theory, as a function of $R$ using the experimentally obtained parameters are plotted as red lines. The measured $r$ that maximizes $p$ are roughly the same as predicted by theory. The dashed parts of the theory lines indicate that a comparison with experiments is not possible in these regions because a fraction of data points at $r \gtrsim 10^7$ \textOmega~is missing due to the finite $V_{ext}$ range used in the experiment.
	}
	\label{fig:Pmaps}
\end{figure*}

\noindent
Determining the maximally obtainable power of a QD in experiment by varying and optimizing $R$ is a cumbersome task. Here we instead show how the power-producing capability of a QD under fixed $\Gamma$, $T$ and $\Delta T$ can be determined by replacing $R$ in the circuit with an external voltage source. During the characterization $I$ is then measured under thermal bias $\Delta T$ as a function of both the external voltage $V_{ext}$ and the gate voltage $V_G$. The data contains information about the power $p=-IV_{ext}$ (denoted by $p$ when $R$ is negligible compared to the resistance of the device\footnote{In our setup $R$ cannot be removed completely, but will always contain at least the resistance of the measurement lines and the pre-amplifier, yielding a minimum $R=14.5\ \text{k}\Omega$. However, the resistance of the device is typically on the order of 1 M$\Omega$ in the most conductive regime and thus $V=V_{ext}-IR\approx V_{ext}$}) and the \textit{effective} resistance of the QD $r=V_{ext}/I$ for given ranges of $V_{ext}$ and $V_G$. The main difference between using a voltage source and using a resistor when studying power production is that the voltage source allows for easy simulation of steady-state load conditions corresponding to changes in $R$ by many orders of magnitude. In addition, the voltage source allows using the device as a heat pump transporting heat from the cold to the hot reservoir since it enables electric work to be performed on the system  ($p<0$). Data obtained using both methods is shown in Fig.~\ref{fig:PVG}. The engine can only produce power ($P>0$) when $V_{ext}=0$, but can both produce ($p>0$) and consume ($p<0$) power when $V_{ext} \neq 0$.
\\
\\
Measuring $p(V_G,V)$ over a range of voltages yields two power-producing regions (Fig.~\ref{fig:Pmaps}a--c). They stem from the two peaks in Figs.~\ref{fig:PnVg}--\ref{fig:PVG} and have different magnitudes due to the different degeneracies of the $N=0$ and the $N=1$ states (see Appendix \ref{app:asym}). The inversion of the thermoelectric signal at $\varepsilon=\mu_c=\mu_h$ (due to electron hole symmetry in the reservoirs) results in the sign change of $V_{ext}$ for the power production regions. The optimal load conditions can be determined by plotting the data in the $p$ and $r$ plane, as shown in Fig.~\ref{fig:Pmaps}d--f (see figure caption for details). A clear maximum in $p$ around $r\approx1-2$ M$\Omega$ is visible, which is consistent with $R_P$ found in an experiment by physically varying the external load resistance for  the same device in Ref.~\onlinecite{josefssonSvilans2018quantum}. This demonstrates that $R_P$ can to a good approximation be found by identifying the effective load that maximizes $p$, which is more convenient than by actually optimizing $R$ physically. 
%
\subsubsection{Theoretical maximum power} 
\begin{figure}[h!] 
	\includegraphics[width=0.42\textwidth]{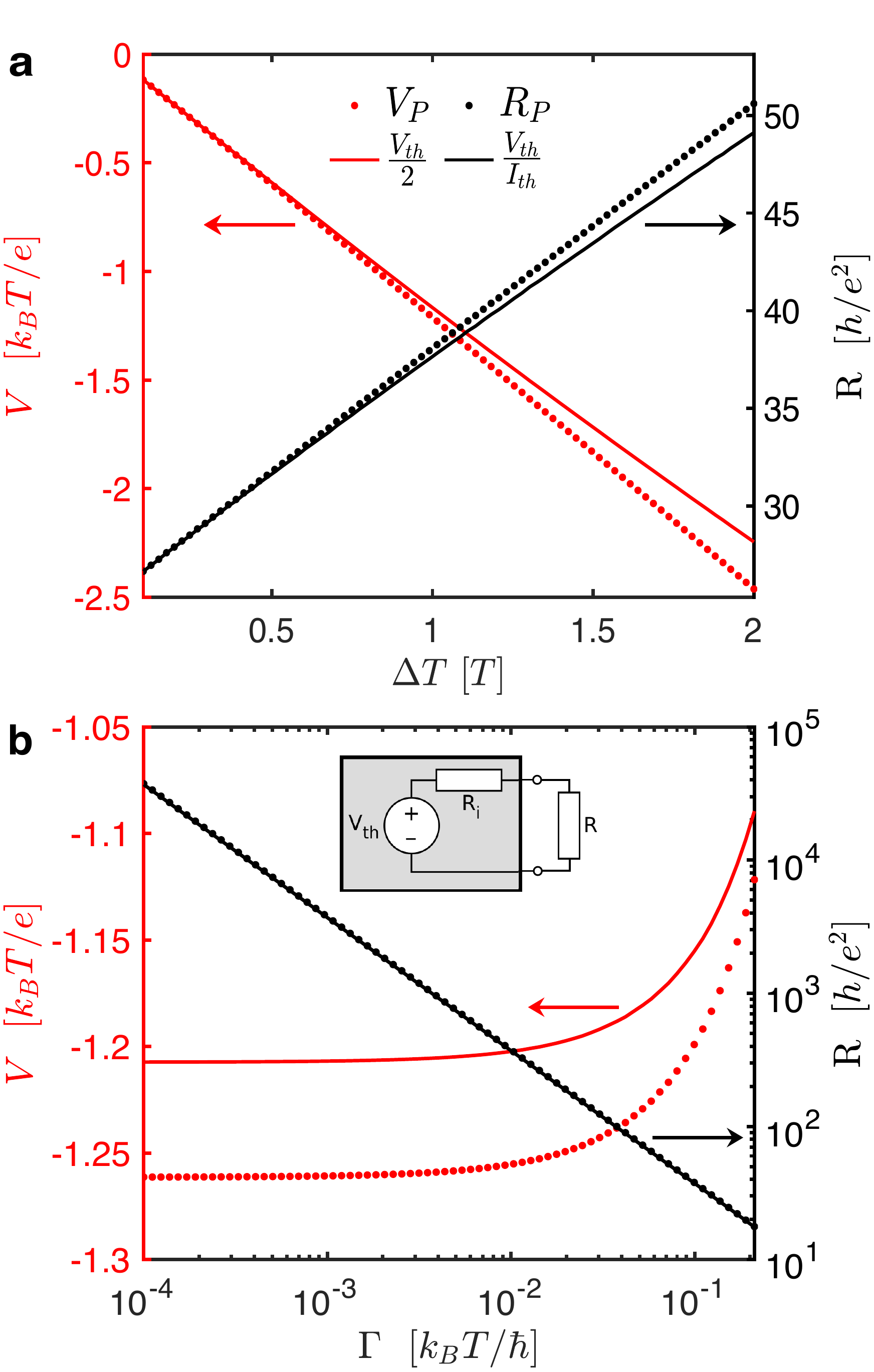}
	\caption{\textbf{Load matching and linear response}. Comparison between the maximum power conditions for the device in the linear response regime (solid lines) and the non-linear regime (dots), both including the second order tunneling effects. In the linear response regime the device can be modelled as a Thevenin equivalent circuit with an ideal voltage source and an internal resistance. The equivalent circuit is inset in \textbf{b} where the grey box describes a QD heat engine. \textbf{b} Voltage and external load at maximum power for for $\hbar \Gamma =0.1k_BT$. \textbf{b} Voltage and external load at maximum power for $\Delta T = T$.}
	\label{fig:Cot_nonlinear}
\end{figure}

\noindent
Since the experimental device used in this study has its $\Gamma$ largely set by the thickness of the of the InP nanowire segments a controlled sweep of the tunnel couplings is not possible. Thus, in this section we will turn to a pure theoretical approach to study how the conditions for maximum power, especially $R_P$, depend on the device parameters. We also investigate the impact of non-linear and second order effects for devices similar to the one used here. For this purpose we solve the Master equations up to second order in $\Gamma$ and calculate maximum $P$ as well as $R_P$ in the linear and non-linear regimes. These results are then compared with the SETA where a linear and non-linear treatment are known to provide comparable results for maximum power.\cite{erdman2017thermoelectric}
\\
\\
First, in order to verify that the second order RTD theory can model the physics at maximum power Figs.~\ref{fig:Pmaps}d--f include theoretical calculations for maximum $P(R)$. These lines agree fairly well with the envelopes of the experimental data, verifying that there is a good agreement also between the measured $p(r)$ and theoretical modeling. There is, however, some discrepancy between the measurements and calculations that deserves a comment. First, the theory slightly underestimates the maximum power. As pointed out in Ref.~\onlinecite{josefssonSvilans2018quantum}, this is most likely due to a small thermoelectric effect in the measurement lines not taken into account by theory. Second, the theory predicts higher $P$ for $R\gtrsim10^7$\textOmega~than the data points show. This is due to the limited $V_{ext}$ range used in the experiment (see Fig.~\ref{fig:Pmaps}a--c) which excludes a fraction of points corresponding to a higher $r$.
\\
\\
Now we turn our attention to investigate the importance of second order and non-linear effects close to maximum power. When a device operates within the linear response regime where $eV/k_BT,\ \Delta T/T\rightarrow 0$ it can be represented as a Thevenin or Norton equivalent circuit (inset in Fig \ref{fig:Cot_nonlinear}). These equivalent circuits consist of an ideal voltage- or current source and an internal load $R_i$. The voltage- and current- sources are characterized by the open circuit voltage and short circuit current of the device, respectively, whereas $R_i$ is characterized by the ratio of the two. In case of a thermoelectric device these quantities correspond to the thermovoltage $V_{th}$ and the thermocurrent $I_{th}$ produced in the linear response regime. From the equivalent circuit one can conclude that the power transferred from a QD to an external load $R$ is maximal when $R=R_i$. The circuit's linear conductance is $G=R_i^{-1}$, and at maximum power production the voltage across the QD is $V_{th}/2$. Hence, 
the power maximized with respect to $V$ can be expressed as $\frac{1}{4}GV_{th}^2$, and as a result $R_P$ is determined from
\begin{equation}
	R_P = I_{th}(V_G^P)/V_{th}(V_G^P), \ \ V_G^P = \max_{V_G}\ GV_{th}^2,
\end{equation}
i.e. $V_G^P$ is the gate voltage that maximizes the linear response power. Within the SETA these quantities can be obtained analytically. We give their expressions below and compare them to results obtained by including non-linear and second order tunneling effects. Formally, the linear response in the SETA assumes $eV/k_BT,\ \Delta T/T,\ \hbar\Gamma/k_BT\rightarrow0$. If $U=\infty$, such that the doubly occupied state is completely unavailable, the current is given by (see appendix \ref{app:linear} for a derivation)
\begin{equation}
	I^{l.r.}=GV+G_T\Delta \tilde{T}=\frac{{e^2\Gamma}}{3k_B\tilde{T}^2}\cdot \frac{\tilde{T}V-\varepsilon\Delta T}{1+\frac{2}{3}\cosh\left( \frac{\varepsilon }{k_B\tilde{T}}\right)+\frac{1}{3}e^{\frac{-\varepsilon}{k_B\tilde{T}}}}.
	\label{eq:linearCurrent}
\end{equation} 
Here, $\tilde{T}$ is the linear response temperature, from here-on taken to be the average temperature $\tilde{T}=(2T+\Delta T)/2$. From Eq. (\ref{eq:linearCurrent}) one can identify the open circuit voltage $V_{th}=\varepsilon\Delta T/\tilde{T}$ and the internal resistance
\begin{equation}
	R_i^{l.r.}=\frac{3k_B\tilde{T}}{e^2\Gamma}\left(1+\frac{2}{3}\cosh\left(\frac{\varepsilon }{k_B\tilde{T}}\right)+\frac{1}{3}e^{\frac{-\varepsilon}{k_B\tilde{T}}}\right).
	\label{eq:linearRP}
\end{equation}
Maximizing $P^{l.r.}=\frac{1}{4}GV_{th}^2$ with respect to $\varepsilon$ yields maximum power production at $\varepsilon\approx 2.533k_BT$, from which the optimal load 
\begin{equation}
 	R^{l.r.}_P\approx 2.507\frac{k_B\tilde{T}}{\hbar\Gamma} \frac{h}{e^2}
	\label{eq:linearRP2.5}
\end{equation}
can be found.
Alternatively, $R_P^{l.r.}$ can be related to the maximum value of the Coulomb peak in the conductance $G_{max}$ (measured at $T$) as 
\begin{equation}
	R_P^{l.r.} \approx\frac{\sqrt{2}}{\sqrt{2}+\frac{4}{3}}\cdot\frac{5.250}{G_{max}} \left(\frac{\Delta T}{2T}+1\right).
	\label{eq:RPlinear2}
\end{equation}
When going beyond the linear response regime it is no longer a trivial task to analytically find the conditions for maximal power production and we will resort to a numerical method. We numerically find $R_P$, and the corresponding voltage output at maximal power $V_P$, using the full non-linear and second order approach and compare it to $I_{th}/V_{th}$ (and $V_{th}/2$) to evaluate the degree of agreement between the linear and non-linear results. In this way we also evaluate how well a Thevenin equivalent circuit can model a realistic QD heat engine. 
\\
\\
Figure~\ref{fig:Cot_nonlinear} demonstrates how the voltage and optimal load scale with $\Delta T$ and $\Gamma$ for the linear and non-linear case. Figure~\ref{fig:Cot_nonlinear}a shows how the deviations between the two cases increase as a function of $\Delta T$, and one can see that the deviations remain relatively small over a wide range of $\Delta T/T$. Figure~\ref{fig:Cot_nonlinear}b shows that the voltages are slightly different, yet remain fairly constant when increasing $\Gamma$ up until $\hbar \Gamma\sim 0.1k_BT$. The results also show that both $R_P$ and $I_{th}/V_{th}$ scale linearly with $\Gamma$ and that there are only minor differences between the two. From this one can also conclude that the maximum power, given by $V_P^2/R_P$, will scale approximately linear with $\Gamma$ as the $\Gamma$-dependence of $R_P$ dominates over that of $V_P$.
\\
\\
In general, the equivalent circuit can be seen as an approximate representation of the device since any deviations remain small over large ranges of $\Delta T$ and $\Gamma$. Hence, even though the internal load is not well defined for a non-linear generator the principle of load matching can still be useful to approximately identify the optimal load of a QD heat engine. To exemplify this we estimate the optimal load for our experimental device using the full non-linear theory, a linear approximation including second order effects $R_i=I_{th}/V_{th}$, and the SETA result in Eq. (\ref{eq:linearRP2.5}). The results are listed in Tab.~\ref{tab:R} and they show that all three methods provide very similar estimates. The relative deviations are of the order of 1\% and will have almost no impact on the output power (see Fig.~\ref{fig:Pmaps}d--f). This means that for all practical considerations $R_P\approx 2.507\frac{k_B\tilde{T}}{\hbar\Gamma} \frac{h}{e^2}$ is a surprisingly good approximation for the load maximizing the power output of a QD heat engine when only a single spin-degenerate orbital contributes to the transport.
\\
\\
The results presented here are valid for QDs with $\hbar\Gamma\ll k_BT$ such that they can be modeled by a perturbative approach up to order $\Gamma^2$. When $\Gamma$ increases to $\hbar\Gamma\sim k_BT$ the QD looses its energy filtering properties.\cite{nakpathomkun2010thermoelectric, kennes2013efficiency} This results in $P$, and therefore also $R_P$, losing their simple dependency on $\Gamma$.
\\
\begin{table}[h]
    \begin{tabular}{| l | l | l | l |}
    \hline
    Measurement & $R_P$ [M$\Omega$] & $V_{th}/I_{th}$ [M$\Omega$] & Eq. (\ref{eq:linearRP2.5}) [M$\Omega$] \\ \hline
    a & 0.940 &  0.935  & 0.930  \\ \hline
    b &  1.157 & 1.150 & 1.144 \\ \hline
    c & 1.438  & 1.430 & 1.430 \\
    \hline    
    \end{tabular}
    \caption{Calculated optimal $R$ for maximal $P$ using the full theory ($R_P$), linear response plus second order tunneling effects ($V_{th}/I_{th}$), and linear response within the SETA from Eq. (\ref{eq:linearRP2.5}). The three cases a--c refer to the experimental conditions defined in the caption of Fig. \ref{fig:Pmaps}a--c, and the calculated optimal $R$s agree well with $r$ for maximum $p$ in Fig. \ref{fig:Pmaps}.}
    \label{tab:R}        
\end{table}
%
\subsection{Efficiency}
\label{subsec:efficiency}
\begin{figure}[ht!]
	\centering
	\includegraphics[width=0.42\textwidth]{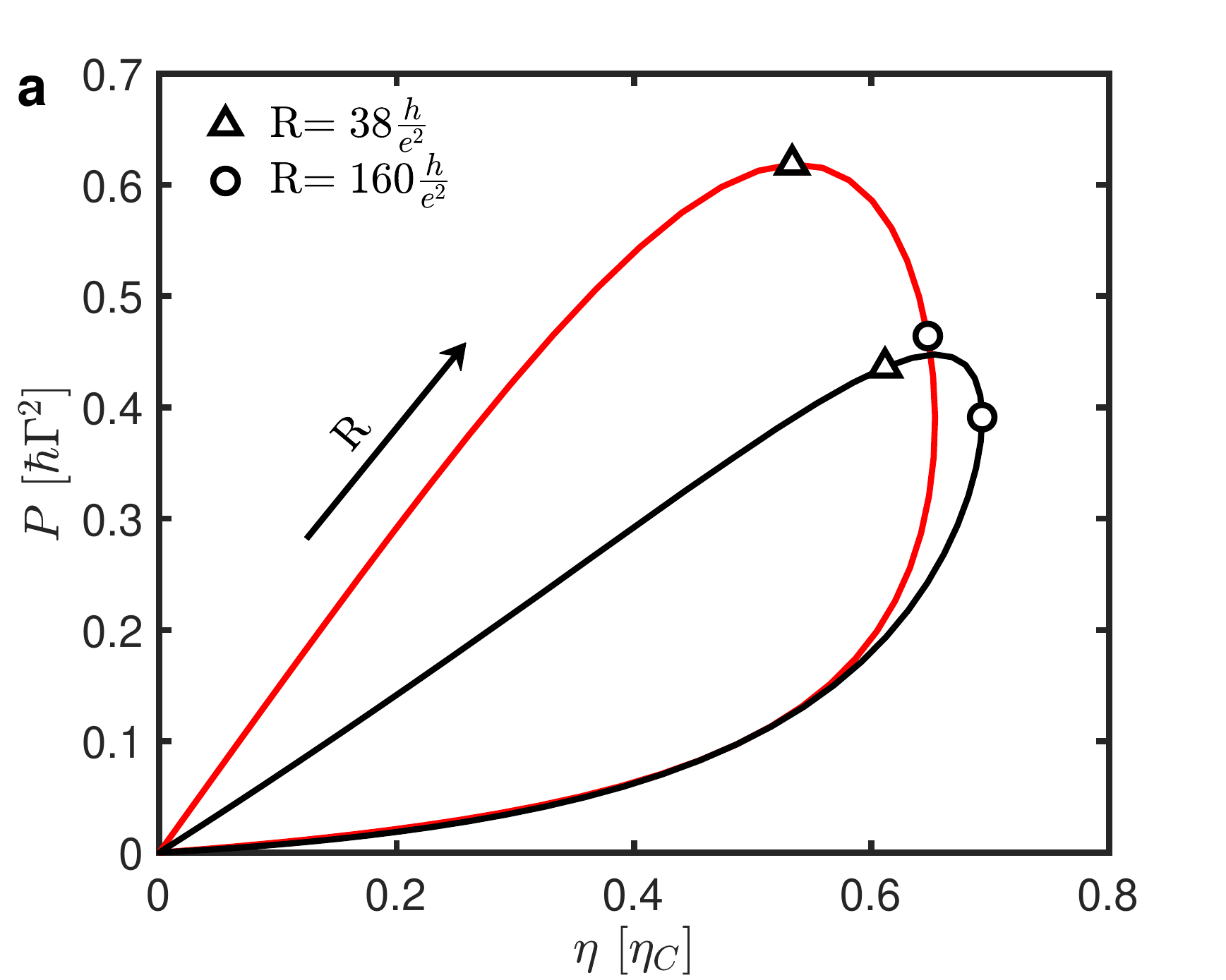}
	\\
	\includegraphics[width=0.42\textwidth]{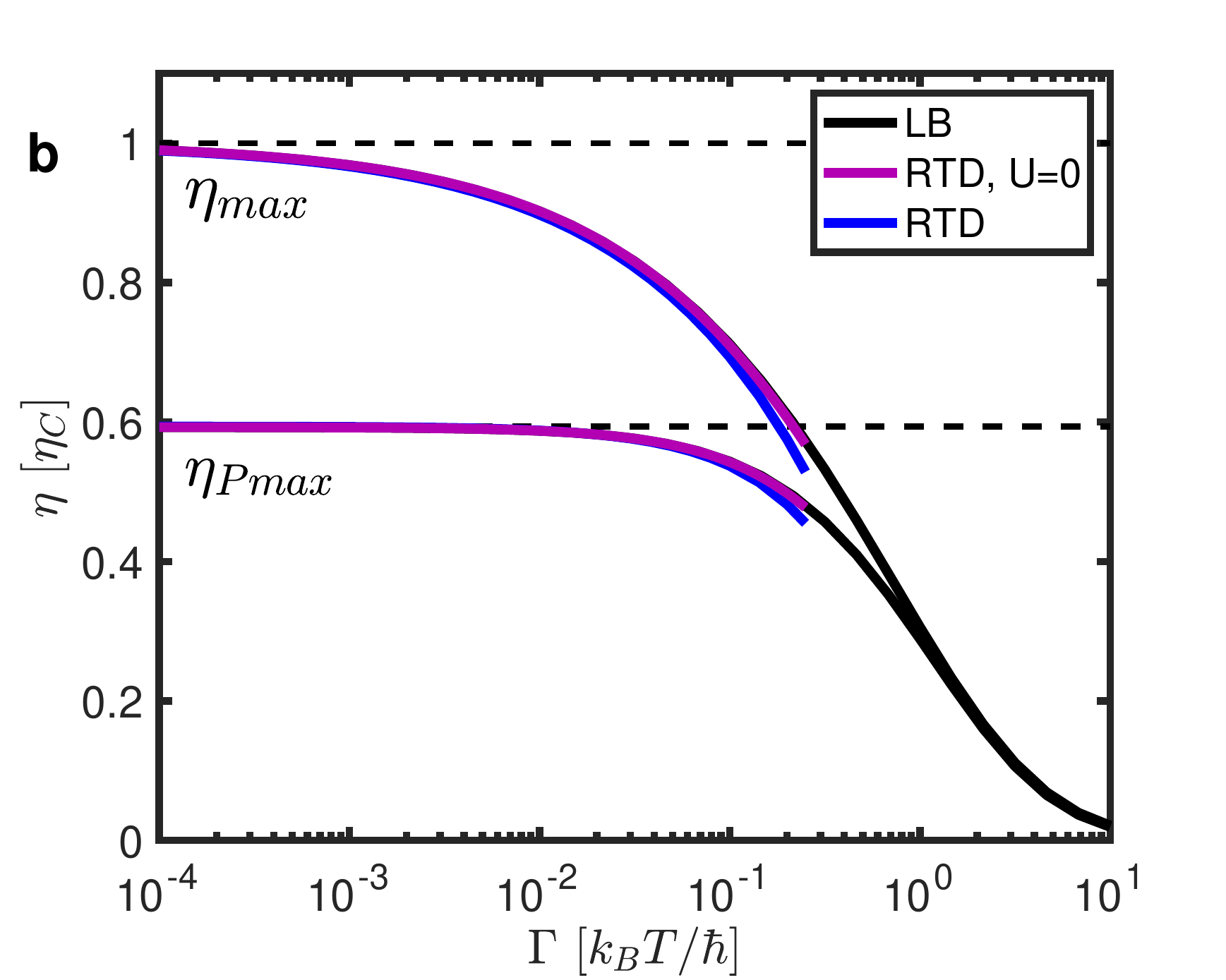}
	\caption{\textbf{Device efficiencies.} \textbf{a} Maximal achievable $P$ and its corresponding $\eta$ as $10^{-2}\frac{h}{e^2}\le R \le 10^{8}\frac{h}{e^2}$ (red line), and $P$ at the maximal achievable $\eta$ for the same $R$-range (black line). 
	The markers indicate $R$ for the engine's highest $P$ (triangles) and $\eta$ (circles). The arrow shows the direction of increasing $R$. 
	Input parameters for the calculations were $\Delta T =T=10\hbar\Gamma/k_B$ and $U=1000\hbar\Gamma$. 
	\textbf{b} Maximum $\eta$ ($\eta_{max}$) and $\eta$ at maximum power ($\eta_{Pmax}$) plotted as a function of $\Gamma$. Input parameters for the calculations were $\Delta T=T$ and $U=1000k_BT$ (blue line) to ensure that the doubly occupied state is unavailable for transport. Calculation results for RTD theory with $U=0$ (purple line) and Landauer-B\"{u}ttiker transport theory (black line) are also included in the plot as references. The dashed lines indicate the SETA result for maximum efficiency $\eta_{max}=\eta_C$ and efficiency at maximum power $\eta_{Pmax}\approx \eta_{CA}=1-\sqrt{T_c/T_h}$, which are independent of $\Gamma$}
	\label{fig:n}	
\end{figure}

\noindent
The previous section explored how to extract the largest amount of power from a QD heat engine. However, the main reason for considering QDs is due to their high $\eta$, which can be optimized in a similar manner as $P$ by tuning $V_G$ and $R$. Figure~\ref{fig:n}a compares the performance of a QD at $V_G$ optimized either for maximum $P$ or maximum $\eta$ as $R$ is being swept, calculated using the full non-linear second order RTD theory. The results show characteristic loop-shaped plots when sweeping $R$ over many orders of magnitude. These plots are suitable for studying the trade-off between $P$ and $\eta$ and they show that for devices similar to the one used in Sec. \ref{sec:RP} there are only small gains in efficiency to be made when optimizing $\eta$ instead of $P$. This means that the conditions for maximum $P$ and maximum $\eta$ are very similar, which is in stark contrast to the results obtained by using the SETA. Within the SETA the QD can obey tight-coupling between the charge and energy flows and operates as a reversible energy converter when carrier populations at $\varepsilon$ are equal in both reservoirs, i.e. $(\varepsilon-\mu_c)/T_c=(\varepsilon-\mu_h)/T_h$.\cite{humphrey2005reversible} During reversible operation the system produces no entropy and there are no net currents, which requires either $R\rightarrow\infty$ or $\varepsilon=\mu_h=\mu_c$. Thus the SETA loop-plots would not close when increasing $R$ to very large values but instead $\eta\rightarrow\eta_C$ as $R\rightarrow\infty$. 
\\
The reason why the loops in Fig.~\ref{fig:n}a close is that the additional second order processes lead to a spread in the energy of the transported electrons around $\varepsilon$. This is because the second order processes involve two coherent tunneling events through population of virtual intermediate states and transport is no longer limited to $\varepsilon$. A consequence of the broadening is that there is a finite heat flow $J_{Q,h}>0$ even when $(\varepsilon-\mu_c)/T_c=(\varepsilon-\mu_h)/T_h$, resulting in $\eta=0$ since $P=0$. This can be observed as a dip in $\eta$ at $\varepsilon=0$ in plots like Fig.~\ref{fig:PnVg}.
\\
\\
The key to a high $\eta$ is to have a QD that couples as weakly as possible to its reservoirs in order to reduce the unwanted heat flow and decrease the deviations from the tight coupling conditions.\cite{nakpathomkun2010thermoelectric} For this reason much theoretical research on QD heat engines assumes $\hbar\Gamma\ll k_BT$ such that the SETA is a valid approximation. However, it is sometimes necessary to include higher order terms even in weakly coupled devices.\cite{josefssonSvilans2018quantum} In order to quantify the impact that the tunnel coupling strength and the second order tunneling processes have on $\eta$, Fig.~\ref{fig:n}b shows how the maximum efficiency $\eta_{max}$ and efficiency at maximum power $\eta_{Pmax}$ scale with $\Gamma$. From the figure it becomes obvious that $\eta_{max}$ shows significant deviations from the SETA-results (dashed lines in Fig. \ref{fig:n}b) even when the tunnel coupling is several orders of magnitude smaller than the thermal energy. $\eta_{Pmax}$ turns out to be less sensitive to increases in $\Gamma$ and retains its weak coupling value over a larger $\Gamma$-range compared to $\eta_{max}$. This is attributed to the fact that maximum efficiency requires voltage conditions where the QD is more Coulomb blockaded compared to maximum power, which results in an exponential suppression of the SETA processes and thus larger influence from the second order contributions. For typical device parameters measured in this study, where the temperatures are on the order of $\Delta T = T\approx10\hbar\Gamma/k_B$, the second order effects are absolutely crucial for proper modeling of the efficiencies. These effects remain important unless $\Gamma$ (and thus $P$) is reduced, or if the temperatures are increased, by several orders of magnitude. However, a practical limitation for increasing the temperatures much further is the experimentally achievable level spacing of the QD orbitals. Once $k_BT$ approaches the level spacing energy, excited states of the QD also participate in the transport, drastically lowering the efficiency.
\\
\\
Figure~\ref{fig:n}b also includes efficiencies calculated for a non-interacting Anderson model, $U=0$, using both the RTD theory and the Landauer-B\"{uttiker} (LB) transport theory (see Appendix B), which can treat the $U=0$ case exactly. The efficiency of this non-interacting system was previously studied in Ref. \onlinecite{nakpathomkun2010thermoelectric} using the LB theory and we include a similar analysis here to verify that a second order perturbative expansion in $\Gamma$ does in fact capture the physics governing the efficiency of a QD heat engine. The two approaches provide almost identical results over the whole validity-range of the RTD theory ($\hbar\Gamma\lesssim 0.25k_BT$), verifying that the RTD theory does capture all important physical effects in this regime. We note that although the two approaches agree well when $U=0$, the LB theory would not be able to model transport through the experimental device used in this study since the electron-electron interactions  are large, and as a consequence the spin-degeneracy does not simply translate to a multiplicative factor of 2.
\section{CONCLUSIONS}
\noindent
We have performed transport calculations using a Master equation approach including tunneling processes up to co-tunneling order together with thermoelectric experiments on a nanowire QD to study the conditions for maximum power production of a QD heat engine where a single level is dominating the transport. In addition, we used the theory to investigate what efficiencies should be obtainable for experimental devices similar to the one studied here.
\\
\\
Generally, it can be a very cumbersome task to experimentally sweep both $V_G$ and $R$ in order to find their optimal values for maximal power production. In order to provide other optimization schemes we have combined theoretical calculations with power measurements to determine the external load that maximizes the power output $R_P$. We showed that $R_P$ can be experimentally identified by measuring the current as a function of both bias and gate voltages without any serial load. $R_P$ is then equal to the value of $V_{ext}/I$ that maximizes $-IV_{ext}$. Furthermore, we theoretically showed that second order tunneling processes and non-linear effects have almost no impact on $R_P$. This means that for experimental implementations of QD heat engines similar to one studied in this paper, $R_P$ can be identified using two of the three quantities $I_{th}\ (\text{short circuit current}),\ V_{th}\ (\text{open circuit voltage}) $ and $G$, which are all straight forward to measure. Alternatively, $R_P$ is well estimated using Eq. (\ref{eq:linearRP2.5}) if $T$, $\Delta T$ and $\Gamma$ are known. The preferred way of identifying $R_P$ hence comes down to what experimental and theoretical tools are available for a specific measurement setup.
\\
\\
We also studied the theoretical maximum efficiency and efficiency at maximum power as a function of both $\Gamma$ and $T$ and found that in contrast to $R_P$, the second order tunnel processes are very important even for $\hbar\Gamma<k_BT$. If one wishes to approach the Carnot efficiency, $\hbar\Gamma/k_BT$ needs to be several orders of magnitude lower than for the experimental device used in this study where $k_BT\approx10-20\hbar\Gamma$. However, the efficiency at maximum power is far less sensitive to these second order effects. The reason for $\eta$ having a strong dependence on $\hbar\Gamma/k_BT$ is that the second order processes lead to an uncertainty in the energy of the transported electrons, and thus an increased heat flow, which effectively decouples the charge current from the heat current.
\section*{ACKNOWLEDGEMENTS}
\noindent
We acknowledge funding from the Swedish Energy Agency (project P38331-1), the Swedish Research Council (projects 621-2012-5122, 2014-5490, 2015-00619 and 2016-03824), the Knut and Alice Wallenberg Foundation (project 2016.0089), and NanoLund. Computational resources were provided by the Swedish National Infrastructure for Computing (SNIC) at LUNARC.
\section*{APPENDICIES}
\label{sec:conclusion}
\appendix
\section{Asymmetric power production}
\label{app:asym}
\noindent
The asymmetric power production around $\varepsilon=0$ in Fig. \ref{fig:PnVg} and $V_G=0.13$ V in Figs. \ref{fig:Pmaps}--\ref{fig:PVG} stems from an asymmetry in the thermoelectrically generated current for different electron numbers on the QD due to spin-degeneracy of the $N=1$ state in combination with a large $U$. This can be understood if one considers the short circuited ($\mu_h=\mu_c=\mu$) version of the heat engine schematic in Fig. \ref{fig:Schemtic}. When $\varepsilon-\mu\sim k_BT$ the most likely electron number on the QD is $N=0$ and the slowest process, which will limit the current, is for an electron to tunnel to the QD from a lead. Since there are two possible transitions for this process, whose matrix elements in the QD sub-space are $\langle\uparrow|H^T|0\rangle$ and $\langle\downarrow|H^T|0\rangle$, this will lead to a high current. In contrast, when $\varepsilon-\mu\sim -k_BT$ the most likely electron number is $N=1$ and the current will be limited by this electron leaving the QD. There is only one possible transition for this to happen, $\langle0|H^T|\downarrow\rangle$ or $\langle0|H^T|\uparrow\rangle$, depending on the spin of the electron on the QD, which will result in a lower current. If we instead considered spin-less electrons or a small charging energy, $U\ll k_BT$, this effect would not be present and it is not captured by theories that do not include electron-electron interactions.
\section{Linear response}
\label{app:linear}
\noindent
In order to derive the linear-response expressions we assume $U=\infty$, $N=0\leftrightarrow1$ and that $\Gamma$ is small enough for the SETA to be accurate. The tunnel couplings to both reservoirs are assumed to be equal $\Gamma=\Gamma_c=\Gamma_h$. The only available states on the QD are then $\{|0\rangle,\ |\uparrow\rangle,\ |\downarrow\rangle\}$, which is further simplified to $\{|0\rangle,\ |1\rangle\}$ by incorporating the spin degeneracy into the tunneling rates.\cite{bonet2002solving} Solving the resulting equation system in Eqs.~(\ref{eq:ME1})--(\ref{eq:ME2}) yields the (non-linear) current
\begin{equation}
	I = e\frac{2\Gamma(e^{x_1}-e^{x_2})}{2e^{x_1+x_2}+3e^{x_1}+3e^{x_2}+4},
\end{equation}
where $x_i = (\varepsilon-\mu_i)/k_BT_i$. The linear conductance ($V=0,\Delta T=0$) is 
\begin{equation} 
	G = \frac{e^2\Gamma}{3k_BT}\cdot \frac{1}{1+\frac{2}{3}\cosh\left(\frac{\varepsilon}{k_BT}\right)+\frac{1}{3}e^{-\frac{\varepsilon}{k_BT}}},
\end{equation}
with $\varepsilon=-e\alpha( V_G-V_G^0)$. Using $G_T=SG$ (where $S=-\varepsilon/eT$ denotes the Seebeck coefficient\cite{turek2002cotunneling}) yields the current in Eq.~(\ref{eq:linearCurrent}). Since the voltage at maximum power is $V_{th}/2$ the optimal gate voltage can be obtained from maximizing $GV_{th}^2$ with respect to $\varepsilon$. Using the fact that the global maximum is located at $\varepsilon>0$ gives a transcendental equation
\begin{equation}
\begin{split} 
	3-\frac{\varepsilon}{k_BT}\sinh\left(\frac{\varepsilon}{k_BT}\right)+2\cosh\left(\frac{\varepsilon}{k_BT}\right)+\\ 
	e^{\frac{-\varepsilon}{k_BT}}\left(1+\frac{\varepsilon}{2k_BT} \right)=0,
\end{split}
\end{equation}
from which $\varepsilon/k_BT\approx2.533$ can be determined numerically. We plug this value into $P=\frac{1}{4}GV^2_{th}$ to obtain the maximally produced power 
\begin{equation}  
	P_{max}=\frac{k_B\Gamma}{T}\cdot\frac{\left(\frac{\beta\Delta T}{2} \right)^2}{1+\frac{2}{3}\cosh\left(\beta\right)+\frac{1}{3}e^{-\beta}}\approx 0.306\frac{k_B\Gamma(\Delta T)^2}{T},
\end{equation}
with $\beta\approx2.533$, which is equivalent to results calculated from a more general formulation in Ref. \onlinecite{erdman2017thermoelectric}.
\\
\section{Landauer-B\"{u}ttiker theory}
\label{app:LB}
\noindent
Landauer-B\"{u}ttiker scattering theory has become a standard tool for modeling electron transport in nano-sized devices where electron-electron interactions are not important, or can be treated within a mean-field approximation. The theory is able to treat the non-interacting QD problem exactly. 
\\
\\
Within this theory the charge and energy currents are given by\cite{landauer1957spatial, buttiker1986four} 
\begin{eqnarray}
	&I_r=-\frac{2e}{h}\int_{-\infty}^{\infty}\tau(\epsilon)\Delta fd\epsilon, \\
	&J_{E,r}=\frac{2}{h}\int_{-\infty}^{\infty}\epsilon\cdot \tau(\epsilon)\Delta fd\epsilon, \\
	\Delta &f=f(\epsilon,\ \mu_h,\ T_h)-f(\epsilon,\ \mu_c,\ T_c).
\end{eqnarray}
Here $\tau(\varepsilon)$ is the transmission function characterizing the system. For a QD with a single energy level at $\varepsilon$ the transmission function is well approximated by
\begin{equation}
	\tau(\epsilon) =\frac{(\hbar\Gamma)^2}{(\hbar\Gamma)^2+(\epsilon-\varepsilon)^2}.
\end{equation}
\bibliographystyle{apsrev_modified}
\bibliography{refs}
\end{document}